\documentclass[iop]{emulateapj}
\usepackage{apjfonts}
\journalinfo{The Astrophysical Journal, 2013, in press}

\shorttitle{A Turbulent Coronal Heating Model for AU Mic}
\shortauthors{Cranmer, Wilner, \& MacGregor}

\begin{document}

\title{Constraining a Model of Turbulent Coronal Heating for
AU Microscopii with X-Ray, Radio, and Millimeter Observations}

\author{Steven R. Cranmer, David J. Wilner,
and Meredith A. MacGregor}
\affil{Harvard-Smithsonian Center for Astrophysics,
60 Garden Street, Cambridge, MA 02138, USA}

\begin{abstract}
Many low-mass pre-main-sequence stars exhibit strong magnetic
activity and coronal X-ray emission.
Even after the primordial accretion disk has been cleared out,
the star's high-energy radiation continues to affect the formation
and evolution of dust, planetesimals, and large planets.
Young stars with debris disks are thus ideal environments for
studying the earliest stages of non-accretion-driven coronae.
In this paper we simulate the corona of AU Mic, a nearby active
M dwarf with an edge-on debris disk.
We apply a self-consistent model of coronal loop heating that was
derived from numerical simulations of solar field-line tangling
and magnetohydrodynamic turbulence.
We also synthesize the modeled star's X-ray luminosity and thermal
radio/millimeter continuum emission.
A realistic set of parameter choices for AU Mic produces
simulated observations that agree with all existing measurements
and upper limits.
This coronal model thus represents an alternative explanation for
a recently discovered ALMA central emission peak that was
suggested to be the result of an inner ``asteroid belt'' within
3 AU of the star.
However, it is also possible that the central 1.3 mm peak is
caused by a combination of active coronal emission and a bright
inner source of dusty debris.
Additional observations of this source's spatial extent and
spectral energy distribution at millimeter and radio wavelengths
will better constrain the relative contributions of the proposed
mechanisms.
\end{abstract}


\section{Introduction}
\label{sec:intro}

Nearly all low-mass stars are believed to have magnetic fields
that influence their surroundings and evolution.
Young stars with ages less than 10--20 Myr exhibit high-energy
activity in the form of hot coronal loops, flares, accretion
shocks, and open-field regions with winds or jet-like outflows
\citep[see, e.g.,][]{FM99,MO07,GN09,Gn13}.
Because the magnetic fields of the star and disk are threaded
together with one another, it is often difficult to disentangle
the contributions from various proposed sources of activity.
On the other hand, the situation may be greatly simplified for
older stars that have evolved past the classical T~Tauri phase.
These stars have lost their dense gas disks, and thus the major
remaining contributor to the star's ultraviolet and X-ray emission
is the presence of magnetic coronal loops.
In that case, the subsequent evolution of a star's {\em coronal
heating} as it starts main sequence hydrogen burning may no longer
involve drastic changes in the source regions, but instead just
be the result of a gradual evolution in parameters related to its
magnetohydrodynamic (MHD) dynamo \citep{HN87,Wr11,St13}.

Pre-main-sequence stars with debris disks are ideal targets for
studying the earliest stages of non-accretion-driven magnetic
activity.
In this paper we focus on AU Microscopii (HD 197481, GJ 803),
a nearby M1Ve flare star with a well-resolved debris disk.
Having an effective temperature of $\sim$3500 K, AU~Mic probably
is not fully convective like the later-type M dwarfs.
Thus, its MHD dynamo may be qualitatively similar to those of more 
massive stars like the Sun.
AU~Mic is bright in X-rays \citep{SS10}, rich in ultraviolet
flaring phenomena \citep{Ro01}, and surrounded by an edge-on
dust disk with a mass of roughly 1 $M_{\rm Moon}$ and a spatial
extent similar to the Kuiper belt in our solar system
\citep[e.g.,][]{Ka04,Li04,AB06,Wd12}.

Recently, \citet{Mc13} observed AU~Mic at millimeter wavelengths
with the Atacama Large Millimeter/submillimeter Array (ALMA).
In addition to the Kuiper-like dust belt, they were able to
distinguish a compact central emission peak with a flux of
$\sim$320 $\mu$Jy.
The origin of this component is not yet known.
The M dwarf's stellar photosphere would generate a blackbody
flux of only about 60 $\mu$Jy at $\lambda = 1.3$ mm, a factor of
5 smaller than the observed emission peak.
In order for a star-sized blackbody to be responsible for
the observed emission, it would need to have a temperature of
roughly 17,000 K.
This kind of dominant chromospheric emission was predicted by
\citet{Hp13} for cool evolved giants, but those stars have
much larger emitting areas and no significant coronal emission.
An M dwarf like AU~Mic may have a thin chromosphere underneath
its hot corona, but its optical depth is not likely to be high
enough to generate a photosphere-like blackbody spectrum.

\citet{Mc13} suggested that the central emission peak of AU~Mic
may be produced by an inner ``asteroid belt'' of cool dust
grains or planetesimals within $\sim$3 AU of the star.
Only about 0.01 $M_{\rm Moon}$ worth of dust material needs to
be present to generate the observed emission peak; this is
comparable to the mass of the asteroid belt in our solar system.
\citet{Mc13} also computed upper limits on the required temperatures
of silicate grains in this proposed belt and found values of
35--75 K.
This range is the same order of magnitude as the expected dust
temperatures at a distance of $\lesssim 3$ AU from an M dwarf.

\begin{figure}
\epsscale{1.05}

\vspace*{0.271in}
\plotone{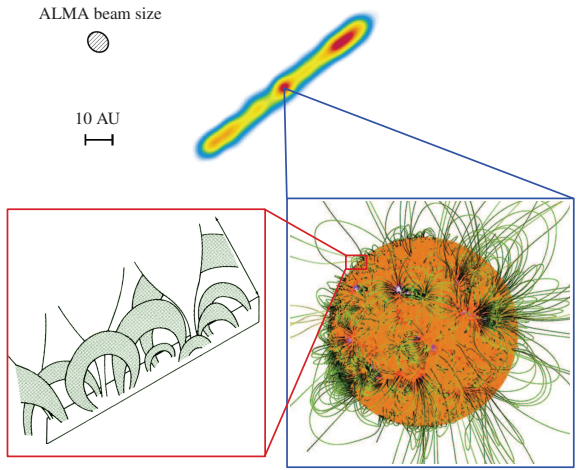}
\caption{Proposed scenario for coronal emission from AU Mic.
Clockwise from top:
ALMA 1.3~mm continuum emission map \citep[adapted from][]{Mc13};
example potential field reconstruction of the solar magnetic
field \citep{Wg12};
schematic illustration of open and closed magnetic regions in
the low corona \citep{Do86}.
\label{fig01}}
\end{figure}

In this paper we propose an alternate explanation for the ALMA
central emission component of AU~Mic.
Since this star has such strong X-ray emission, we explored the
possibility that a collection of {\em hot coronal loops} on its
surface could be responsible for similarly strong thermal emission
at 1.3 mm.
Figure \ref{fig01} illustrates the suggested MHD circumstellar
environment, where magnetic loops with a continuous distribution
of sizes are assumed to fill the corona with $\sim 10^{6}$ K plasma.
We aim to create a single model of these loops that reproduces
the available observations at X-ray and
millimeter wavelengths.
\citet{Wh94} also explored the use of radio continuum measurements
as constraints to models of M dwarf coronal heating.
In a way, this paper follows on from the general ideas described
by \citet{Wh94}, but our models are based on self-consistent
physical sources of plasma heating inspired by recent advances
in understanding the Sun's well-resolved corona.

In Section~\ref{sec:model} of this paper, we describe the model
of coronal heating that we apply to AU~Mic.
Section~\ref{sec:radio} summarizes how we synthesize observable
quantities in the X-ray, radio, and millimeter wavelength bands
for later comparison with observations.
In Section~\ref{sec:results} we present the computed coronal loop
parameters for AU~Mic as well as synthetic X-ray luminosities
and radio/mm spectra.
Section \ref{sec:conc} concludes the paper with a brief summary
of the main results and suggestions for future improvements.

\section{Coronal Heating Model}
\label{sec:model}

In this section we describe a model of turbulence-driven coronal
activity motivated by the present-day Sun.
The model adopts a time-steady description of coronal heating and
does not attempt to simulate the full range of intermittent and
chaotic MHD dynamics that is evident in high-resolution solar
images \citep[e.g.,][]{Ci13}.
Nonetheless, such a phenomenological description of turbulence has
been shown to accurately reproduce many properties of coronal heating
and wind acceleration for the Sun \citep{Cr12}.
Also, \citet{Cr09} showed how this model performs well in predicting
the X-ray activity of young stars undergoing active accretion, and
\citet{CS11} used the model to compute the mass-loss rates
of stellar winds for stars with a wide range of ages and masses.

The basic idea is that magnetic field lines at the stellar
photosphere are continually jostled by a stochastic source of
mechanical energy, and this gives rise to MHD waves that propagate
up to larger heights and eventually dissipate.
In tandem with wave generation, the random footpoint motions
produce an increase in the overall magnetic energy via the twisting,
shear, and braiding of field lines \citep[e.g.,][]{P72}.
The transport and dissipation of magnetic energy from both waves
and braiding can be described using the unifying language of
{\em turbulent cascade} \citep{vB86,Go00,Rp08,vB11}.
The presence of radiative cooling in such a system produces a thermal
instability that often leads to the coexistence of a cool
($T \lesssim 10^{4}$ K) chromosphere and a hot
($T \gtrsim 10^{6}$ K) corona.
In the ionized corona, heat conduction also makes for a ``thermostat''
effect in which the time-steady spatial distribution of $T$ is
much smoother than the spatial distribution of the heating rate.

There are several possible sources of stochastic mechanical energy
at the photospheric base.
Convective granulation surely plays a major role for most low-mass
stars \citep{Mz04}.
For stars undergoing active accretion, an additional source of
energy can exist in the form of ripples from the impact of gas
clumps falling down from above \citep{Cr08}.
It has also been suggested that the presence of planets in a
magnetized stellar wind may give rise to MHD fluctuations that
propagate back down to the star \citep[e.g.,][]{Lz12}.
For the purposes of this paper, we will treat the total available
mechanical energy flux at the photosphere as a free parameter.

We describe the star itself by its fundamental parameters (mass,
radius, luminosity, and metallicity) and ignore rotation.
The dynamics of the jostled magnetic flux tubes are described
by three additional parameters defined at the photosphere:
mass density $\rho_{\ast}$, magnetic field strength $B_{\ast}$, and
the mean energy flux $F_{\rm A}$ of Alfv\'{e}n waves that
propagate upwards as a result of the stochastic motion.
To compute the density as a function of effective temperature
$T_{\rm eff}$ and gravity $g$, we used the standard criterion
that the Rosseland mean optical depth $\tau_{\rm R}$ should have
a value of $2/3$ in the photosphere, which demands
\begin{equation}
  \tau_{\rm R} \, \approx \, \kappa_{\rm R} \rho_{\ast} H_{\ast}
  \, = \, 2/3 \,\, .
  \label{eq:tauR}
\end{equation}
The Rosseland mean opacity $\kappa_{\rm R}$ was interpolated from
tables given by \citet{Fg05}, and we defined the density scale
height in the photosphere as
\begin{equation}
  H_{\ast} \, = \, \frac{k_{\rm B} T_{\rm eff}}{\mu m_{\rm H} g}
  \,\, ,
\end{equation}
where $k_{\rm B}$ is Boltzmann's constant and $m_{\rm H}$ is
the mass of a hydrogen atom.
\citet{CS11} found that, for cool-star photospheres, the mean
atomic weight $\mu$ is a relatively slowly varying function of
$T_{\rm eff}$; we used their parameterization.

To specify the photospheric magnetic field strength, we assumed
that $B_{\ast}$ remains linearly proportional to the so-called
equipartition field strength (i.e., the value at which magnetic
pressure balances gas pressure).
We used the proportionality constant found from observations by
\citet{CS11} to specify
\begin{equation}
  B_{\ast} \, = \, 1.13 \,
  \sqrt{\frac{8\pi \rho_{\ast} k_{\rm B} T_{\rm eff}}{\mu m_{\rm H}}}
  \label{eq:Beq}
\end{equation}
for the footpoints of coronal loops.
The Alfv\'{e}n wave energy flux can be written in terms of the
surface velocity amplitude $v_{\perp}$ of transverse MHD waves, with
$F_{\rm A} = \rho_{\ast} v_{\perp}^{2} V_{\rm A}$, and the
Alfv\'{e}n speed is given by
$V_{\rm A} = B_{\ast} / \sqrt{4\pi\rho_{\ast}}$.
We can make an initial estimate of the expected velocity amplitude
using turbulent convection models.
\citet{CS11} used the models of \citet{Mz02} for main sequence and
evolved stars.
We find below, however, that for AU Mic the \citet{Mz02} models
predict an amplitude that is far too small to explain its observed
coronal heating and X-ray activity.
(Our understanding of the magnetic dynamos of young, rapidly rotating
stars is still incomplete; see Section \ref{sec:conc}.)

In general, we assume that magnetic flux tubes fill only a fraction
$f_{\ast}$ of the stellar photosphere, so that the mean magnetic
flux density of the star would be given by
$\langle B \rangle \approx f_{\ast} B_{\ast}$.
For the Sun, $f_{\ast}$ varies between about $10^{-3}$ and $10^{-1}$
over location and activity cycle \citep{SH89}.
The local magnetic field strength $B(r)$ inside a flux tube
drops rapidly from $B_{\ast}$ to $\langle B \rangle$ as a function
of increasing height.
However, for young stars in the ``saturated'' part of the
age--activity--rotation diagram, \citet{Sa01} and \citet{CS11}
found that $f_{\ast} = 1$ is not a bad approximation to make.
We will assume this for AU~Mic as well (see additional discussion
in Section \ref{sec:results}), and this also allows us to
assume that $B_{\ast}$ remains constant over each coronal loop.

We note that $f_{\ast}$ describes the filling factor of all magnetic 
flux tubes, of both polarities, that pass through the stellar surface.
By itself, our assumption of $f_{\ast} = 1$ does not directly imply a
prediction for the fraction of the magnetic field lines that are open
(presumably to the stellar wind) versus those that are are closed.
The open/closed fraction depends on both the overall level of gas
pressure in the stellar wind and the spatial patterns of imbalance
between flux tubes with positive and negative polarities
\citep{Cl03,CvB10}.
Below, we do assume the entire surface is covered by loops of varying
lengths, but the coronal heating---and thus the X-ray and radio/mm
emission---from the longest loops is likely to be not too different
from the emission from open-field regions \citep[see, e.g.,][]{Sj04}.

To specify the rate of plasma heating, we use a general expression
for the rate of energy flux in the cascade from large to small
eddies that was derived from analytic and numerical studies of
MHD turbulence.
Following \citet{Cr09}, we define the heating rate (in units of
power generated per unit volume) as
\begin{equation}
  Q \, = \, \frac{\rho v_{\perp}^3}{\lambda_{\perp}}
  \left( \frac{\lambda_{\perp} V_{\rm A}}{v_{\perp} L}
  \right)^{\alpha} \, ,
  \,\,\,\,\, \mbox{where} \,\,\,\,\,
  \alpha \, = \, \frac{2 + 420 {\cal R}}{1 + 280 {\cal R}}
  \label{eq:Qturb}
\end{equation}
and ${\cal R} = v_{\perp} / V_{\rm A}$.
The quantity $\lambda_{\perp}$ is a transverse correlation length
for the largest driving eddies, and is typically found to be
of the same order of magnitude as the radius of the flux tube
\citep[e.g.,][]{Ho86}.
By convention, $L$ is the half-length of the coronal loop.
Nonzero values of the exponent $\alpha$ describe departures from
an ideal \citet{K41} hydrodynamic cascade; such departures occur
because of the specific MHD nature of wave-packet collisions that
comprise the eddies in the presence of a strong background field.
In closed loops, $\alpha$ can vary between 1.5 and 2.
Our expression for the dependence of $\alpha$ on the wave
amplitude $v_{\perp}$ and Alfv\'{e}n speed $V_{\rm A}$ was derived
from the models of \citet{Rp08}.

Even though Equation (\ref{eq:Qturb}) utilizes constant values
for $v_{\perp}$, $\lambda_{\perp}$, $L$, and $B_{\ast}$ (in the
definition of $V_{\rm A}$), the density $\rho$ is known to
decrease rapidly as a function of height in a stellar atmosphere.
Thus, because $V_{\rm A} \propto \rho^{-1/2}$, the heating rate
itself varies with density as $Q \propto \rho^{1 - (\alpha / 2)}$.
For $\alpha = 2$ the heating rate is constant
\citep[as was assumed by][]{RTV}, and for $\alpha < 2$ the heating
is stronger at the high-density footpoints.
The values of $\rho$ and $V_{\rm A}$ at the ``coronal base'' (i.e.,
just above the sharp transition region between chromosphere and
corona) are not the same as the corresponding values in the
photosphere; these quantities are computed self-consistently by the
model and are not specified as inputs.

In order to determine the time-steady coronal response to a given
heating rate $Q$, we must consider how the injected heat is
transported along the field by conduction and lost by radiation.
We use the energy balance model of \citet{Ma10} to solve for
the temperature $T$ and electron number density $n_e$
as a function of distance along a given loop of half-length $L$
\citep[see also][]{RTV,AS02}.
These quantities are specified by first computing the peak
temperature $T_{\rm max}$ at the top of the loop and the
base pressure $P$.
\citet{Ma10} parameterized the heating rate as $Q = h T^{a} P^{b}$,
where $h$ and $P$ are assumed to be constants as a function of
distance.
For the heating model of Equation (\ref{eq:Qturb}), the
exponents are given by $a = -b = (\alpha / 2) - 1$.
\citet{Ma10} derived two scaling laws that allow one to solve for
any two of the four loop quantities ($L$, $T_{\rm max}$, $P$, $h$)
if the other two are provided.
In our case, these scaling laws are given by
\begin{equation}
  T_{\rm max} \propto
  h^{2/(2\alpha + 3)} L^{(2 + \alpha)/(2\alpha + 3)}
  \,\, , \,\,\,\,
  P \propto T_{\rm max}^{3} / L
  \,\, ,
  \label{eq:scale}
\end{equation}
and the unspecified constants of proportionality are described
in detail by \citet{Ma10}.
These constants depend on the exponent $\alpha$, the heat
conductivity coefficient $\kappa = 1.1 \times 10^{-6}$
erg cm$^{-1}$ s$^{-1}$ K$^{-7/2}$, and the assumed properties
of radiative cooling.
For the latter, we assumed $Q_{\rm cool} = \chi P^{2} T^{-2.5}$,
with a value of $\chi = 3.65 \times 10^{12}$ cm$^4$ s$^{-1}$ K$^{3/2}$,
which we enhanced slightly over the solar value because of the higher
metallicity of AU~Mic \citep[see Equation 21 of][]{CS11}.

In order to model the thermal state of a given loop with half-length
$L$, we use Equation (\ref{eq:Qturb}) to determine $h$ in terms of
the other input parameters, and we use the scaling laws given in
Equation (\ref{eq:scale}) to solve for $T_{\rm max}$ and $P$.
The density at the coronal base is determined mainly by $P$, but
it needs to be specified initially in order to compute $V_{\rm A}$
and $\alpha$ in Equation (\ref{eq:Qturb}).
Thus, we solve for these quantities iteratively by starting with
the initial guess that the density is given by $\rho_{\ast}$.
Cycling back through the equations gives rise to a converged and
self-consistent value of $\rho$ (and thus also $V_{\rm A}$ and
$\alpha$) typically within 10--15 iterations, and we run it for
a total of 50 iterations to ensure accuracy.
This converged value of the coronal base density is often
3--6 orders of magnitude smaller than $\rho_{\ast}$, which
produces typical basal $n_e$ values of $10^{10}$--$10^{14}$
cm$^{-3}$.

Once $T_{\rm max}$ and $P$ are known for a given loop, we solved for
the spatial dependence of temperature and density along the loop
length $s$.
\citet{Ma10} found that the temperature dependence $T(s)$ can be
written as the inverse of an incomplete beta function; we used
Equation (25) of \citet{Ma10} to obtain $T(s)$, and then we
combined it with the ideal gas equation of state to solve for
$n_{e}(s)$ as a function of $P$ and $T(s)$.
The minimum value of $T$ at the coronal base was fixed at $10^4$ K.

Lastly, we must take account of the fact that there should be a
continuous distribution of loop lengths $L$ across the surface of
the star.
We specify a normalized probability distribution $N(L)$ that
describes the chances of finding a loop at any given value of $L$
at a random location on the star.
The shape of $N(L)$ depends on many details about the strength and
topology of the magnetic field.
The large-scale magnetic geometry of M dwarfs is beginning to be
understood observationally \citep[e.g.,][]{Gs13}, but we know from
the Sun that the X-ray and UV emission is dominated by activity on
spatial scales much smaller than have been resolved so far on
other stars.
Thus, we follow \citet{Cr09} and assume a power-law probability
distribution, $N(L) \propto L^{-\varepsilon}$, with sharp cutoffs
below minimum and maximum loop lengths $L_{\rm min}$ and
$L_{\rm max}$.
The exponent $\varepsilon$ in the loop-length distribution is
another key free parameter of this model.
The shortest loops are defined geometrically as those for which
the central hole in the torus shrinks to zero; i.e.,
$L_{\rm min} = \pi r_{\perp} / 2$, where $r_{\perp}$ is the
modeled poloidal radius of the loop (see Section \ref{sec:results}).
The longest loops are assumed to have lengths of order the stellar
radius.
We note that some young stars may have loops that extend
out to even larger heights \citep{Jd05,Aa12}.
Nonetheless, it is not known what fraction of the very longest
loops remain stable and closed in the presence of a stellar wind.
Thus, we set $L_{\rm max} = R_{\ast}$ and assume that the coronal
emission is relatively insensitive to the details of what happens
for the small fraction of the star covered by the longest loops.

\section{Synthesis of X-Ray and Radio/Millimeter Emission}
\label{sec:radio}

To simulate a full distribution of loops covering the surface of
AU~Mic, we constructed 100 coronal models on a logarithmic grid in
$L$ that spans the $\sim$3.5 orders of magnitude between
$L_{\rm min}$ and $L_{\rm max}$.
We synthesized the observable quantities described below for the
100 models individually---assuming for each that the star is filled
with loops of a given length---then we convolved them together
using $N(L)$ to produce a result that is weighted properly over the
statistical distribution of loop sizes.

We computed the total X-ray luminosity $L_{\rm X}$ under the
optically thin assumption that all radiation escapes from the
emitting regions.
To maintain continuity with past observations of cool-star
X-rays, we chose to use the response function of the {\em ROSAT}
Position Sensitive Proportional Counter (PSPC) given by \citet{Ju03}.
This function has nonzero sensitivity between about 0.1 and 2.4 keV,
with a minimum around 0.3 keV that separates the hard and soft bands.
\citet{Cr09} presented the temperature-dependent radiative loss rate
$\Lambda_{\rm X}(T)$ that is consistent with this response function,
which is then used to estimate the X-ray luminosity,
\begin{equation}
  L_{\rm X} \, = \, 4\pi R_{\ast}^{2} \int dz \,\,
  [ n_{e}(z)^{2} ] \, \Lambda_{\rm X}[T(z)]  \,\, .
\end{equation}
We integrated down through the length of each loop, from $z=L$ to
$z=0$, under the simplifying geometric assumption that the loop
is oriented vertically.
In other words, we assumed that each loop has been ``snipped in two''
and the two ends are pointing radially upward.
We also ignored foreshortening effects that would alter the
path lengths through loops close to the stellar limb.

Calculating the emission at radio and millimeter wavelengths was
slightly more complicated than the X-ray luminosity because we can
no longer assume an optically thin emitting region.
Thus, we solved the diffuse term in the formal solution to the
equation of radiative transfer by integrating both the optical
depth $\tau_{\nu}$ and the specific intensity $I_{\nu}$
simultaneously, with
\begin{equation}
  I_{\nu} \, = \, \int dz \,\, \chi_{\nu} \,
  \frac{2k_{\rm B}T \nu^2}{c^2} \, e^{-\tau_{\nu}}
\end{equation}
\begin{equation}
  \tau_{\nu} \, = \, \int dz \,\, \chi_{\nu}
\end{equation}
and we made the standard assumption that the source function is
given by the Rayleigh-Jeans tail of the local Planck function.
Because the plasma conditions vary rapidly as a function of
position $z$ along the loop, we did {\em not} make the other
standard assumption that the source function is constant over
the emitting area.
As above, we integrated down from $z=L$ to $z=0$, so that the
resulting optical depth increases as $z$ decreases.

For wavelengths $\lambda$ between 0.01 and 100 cm, we assumed the
radiation is dominated by thermal free-free emission (bremsstrahlung)
and that the opacity is given by
\begin{equation}
  \chi_{\nu} = 0.01 \, n_{e}^{2} T^{-3/2} \nu^{-2} \,
  \ln \left( \frac{4.7 \times 10^{10} \, T}{\nu} \right)
  \label{eq:freefree}
\end{equation}
in cgs units \citep[see][]{Dk85,Gu02}.
For the Gaunt factor (i.e., the natural logarithm term above),
we used a fully ionized approximation that should apply for
$T \gtrsim 3 \times 10^{5}$ K.
Equation (\ref{eq:freefree}) also assumes $\nu \gg \nu_{p}$,
where $\nu_p$ is plasma frequency in Hz; this condition is
satisfied easily for the ALMA submillimeter wavelengths of
interest and is satisfied marginally for longer radio wavelengths.
For a star at distance $D$, we converted the surface-averaged
specific intensity $I_{\nu}$ to flux $S_{\nu}$ by assuming that short
loops (with $L \ll R_{\ast}$) generally dominate the emission and
that there is no limb brightening or limb darkening.
Thus, we used $S_{\nu} = \pi R_{\ast}^{2} I_{\nu} / D^{2}$.

\section{Results for AU Mic}
\label{sec:results}

Table 1 gives the basic stellar parameters that we used
either as inputs (upper part) or as after-the-fact constraints
(lower part) on the models.
The photospheric density $\rho_{\ast}$ and magnetic field strength
$B_{\ast}$ were computed using the equations discussed above.
The photospheric turbulence length scale $\lambda_{\perp}$ and
flux tube radius $r_{\perp}$ were scaled down from canonical
solar values of 300 km and 200 km, respectively, by the ratio of
photospheric scale heights.
The solar value of $\lambda_{\perp}$ comes from models of
Alfv\'{e}n wave damping in the fast solar wind \citep{CvB05} and
$r_{\perp}$ comes from measurements of magnetic bright points
\citep{Bg95,Bg07}; both are related closely to the horizontal
sizes of magnetic features sitting between the granulation cells.
Recent convection models \citep[e.g.,][]{Rf04,Mg13} show that the
diameters of granules remain linearly proportional to the
vertical scale height $H_{\ast}$ over a wide range of stellar
parameters.\footnote{%
Note, however, that rotation is {\em not} one of the stellar
parameters varied in most stellar convection simulations.
See Section \ref{sec:conc} for additional discussion of how rapid
rotation may invalidate some traditional ideas about convection
and magnetic activity.}
We assume that intergranular features like magnetic flux tubes
scale with $H_{\ast}$ similarly as the cells.
With the values given above, we found that the coronal loops of
AU~Mic span about 3.5 orders of magnitude in length between
$L_{\rm min} = 190$ km and $L_{\rm max} = 5.83 \times 10^5$ km.
\begin{deluxetable}{lll}
\tablecaption{AU Mic Stellar Parameters}
\label{table01}
\tablewidth{0pt}
\startdata
\hline
Distance      & 9.91 pc            & \citet{vL07}\\
$M_{\ast}$    & 0.5 $M_{\odot}$    & \citet{SC06}\\
$R_{\ast}$    & 0.838 $R_{\odot}$  & \citet{Ho09}\\
$T_{\rm eff}$ & 3493 K             & \citet{Ho09}\\
$\mbox{[Fe/H]}$ & 0.154              & \citet{Ho09}\\
$\rho_{\ast}$ & $9.2 \times 10^{-7}$ g cm$^{-3}$ & Equation (\ref{eq:tauR})\\
$B_{\ast}$    & 2200 G             & Equation (\ref{eq:Beq})\\
$\lambda_{\perp}$ & 182 km         & see text \\
$r_{\perp}$      & 121 km         & see text \\
\hline
$L_{\rm X}$   & $5.549 \times 10^{29}$ erg s$^{-1}$ & \citet{Hu99}\\
$S_{\nu}$ ($\lambda = 1.3$ mm) & 320 $\pm$ 60 $\mu$Jy &
  \citet{Mc13}\tablenotemark{a} \\
$S_{\nu}$ ($\lambda = 2.0$ cm) & $<$ 210 $\mu$Jy & \citet{Wh94} \\
$S_{\nu}$ ($\lambda = 3.7$ cm) & $<$ 120 $\mu$Jy & \citet{Wh94} \\
\enddata
\tablenotetext{a}{AU~Mic was observed by ALMA with its Band 6
receivers over four 2~hr long scheduling blocks between 2012 April 23
and 2012 June 16.  These data are designated
ADS/JAO.ALMA\#2011.0.00142.S.}
\end{deluxetable}

For AU~Mic, we assumed the photospheric magnetic filling factor
$f_{\ast}$ is equal to 1.
AU~Mic appears to be safely inside the ``saturated'' regime of
stellar activity that is consistent with this assumption.
Its rotation period $P_{\rm rot} \approx 4.8$ days implies a
dimensionless Rossby number (i.e., $P_{\rm rot} / \tau_{c}$, where
$\tau_c$ is the convective turnover time) of 0.05--0.1 \citep{Hb07}.
Figure 7(b) of \citet{CS11} shows $f_{\ast} \approx 1$ for a large
sample of cool stars in this range of Rossby number.
AU~Mic also has an X-ray luminosity ratio $L_{\rm X}/L_{\rm bol}$ of
0.0015, which is in the saturated part of the empirical
age--activity--rotation diagram \citep[e.g.,][]{Wr11}.
\citet{Te04} estimated the filling factor of ``solar-like active
region'' plasma in the corona of AU~Mic to be roughly 0.9--1.
Although this measurement does not constrain the {\em photospheric}
value of $f_{\ast}$, it helps to show that the coronal flux tubes
eventually fill the entire circumstellar volume.

The one remaining major parameter that is not yet determined for
AU~Mic is the surface flux of Alfv\'{e}n waves.
\citet{CS11} used the models of \citet{Mz02} to produce a
fitting formula that gives this surface flux as a function
of a star's $T_{\rm eff}$ and $\log g$.
For the parameters of AU~Mic this formula gives 
$F_{\rm A} \approx 2 \times 10^{6}$ erg cm$^{-2}$ s$^{-1}$, which
is equivalent to $v_{\perp} \approx 0.02$ km s$^{-1}$.
On the other hand, both observations of microturbulent broadening
\citep{Gm82} and three-dimensional convection simulations
\citep{Wd09,Bk11} show vigorous overturning motions with velocities
of order 1--2 km s$^{-1}$ for M dwarfs.
For the Sun, the granulation velocities and surface wave amplitudes
are roughly of the same order of magnitude as one another, but we do
not know if this holds true for M dwarfs.
Thus, we treat $v_{\perp}$ as a free parameter, but we also note
that values of order 1--2 km s$^{-1}$ may be the most realistic.

\begin{figure}
\epsscale{1.10}
\plotone{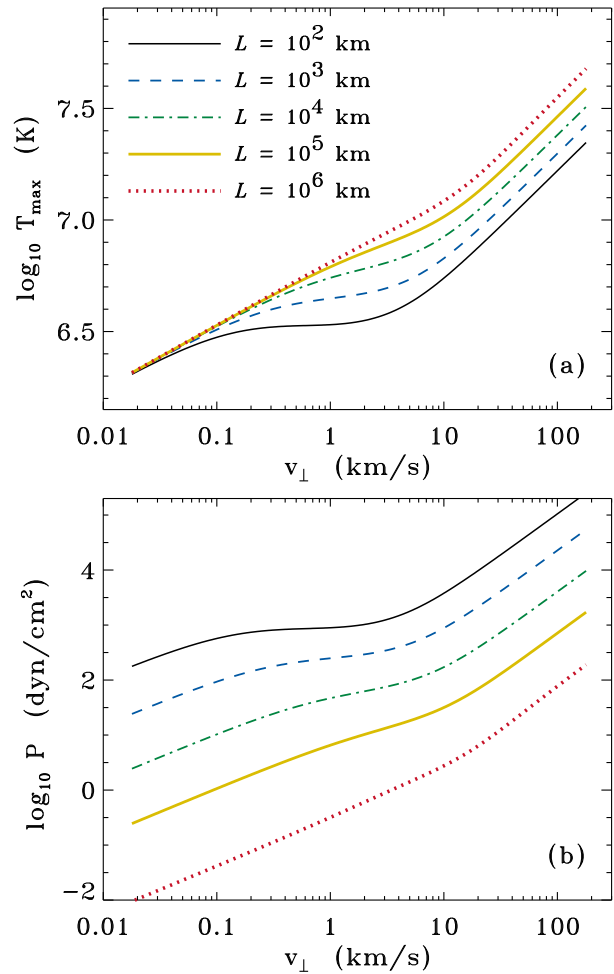}
\caption{Dependence of (a) loop-top coronal temperature and
(b) gas pressure at the coronal base on footpoint wave amplitude
$v_{\perp}$ (horizontal axis) and loop length $L$ (see caption for
curve identification) for our standard coronal model of AU~Mic.
\label{fig02}}
\end{figure}
We computed models of coronal heating for a grid of loops having
lengths $L$ between $10^2$ and $10^6$ km, and footpoint wave amplitudes
$v_{\perp}$ between 0.02 and 200 km s$^{-1}$.
Figure \ref{fig02} shows the resulting dependence of $T_{\rm max}$
and $P$ on these two parameters.
Each of these models had its own iterated value of the $\alpha$
exponent, and the largest values ($\alpha \approx 2$) tended to
occur for the lowest velocity amplitudes and the longest loops;
the smallest values ($\alpha \approx 1.5$) occurred for the
largest amplitudes and shortest loops.
The overall dependence of $T_{\rm max}$ and $P$ on the two varied
parameters is close to what is expected from the scaling laws given
in Section \ref{sec:model}.
For example, if we use the average value of $\alpha = 1.78$ for
this set of models, the scaling laws give
\begin{equation}
  T_{\rm max} \propto v_{\perp}^{0.37} L^{0.034} 
  \,\, , \,\,\,\,
  P \propto v_{\perp}^{1.12} L^{-0.90}
\end{equation}
which agrees well with the plotted results.
The curves shown in Figure~\ref{fig02} are not exact power laws
because $\alpha$ itself is not constant.

In order to begin the process of testing our coronal models against
real observations of AU~Mic, we assembled together the temperature
and density distributions for individual loop lengths into
surface-averaged models using the power-law length distribution
$N(L) \propto L^{-\varepsilon}$ discussed above.
Various measurements of solar features have constrained the value
of $\varepsilon$ to be of order 2--2.5 \citep{As00,As08,Cl03},
and the \citet{Cr09} study of T~Tauri star X-ray emission also
found that $\varepsilon \approx 2.5$ produced the most reasonable
results.
For completeness, we varied $\varepsilon$ between 0 and 2.5,
noting that values of the exponent greater than 2.5 produce
distributions that are only marginally different from those with
$\varepsilon \approx 2.5$ (i.e., once it is peaked sharply at
the shortest length scales, making it even steeper does not change
the resulting weighting significantly).

\begin{figure}
\epsscale{1.10}
\plotone{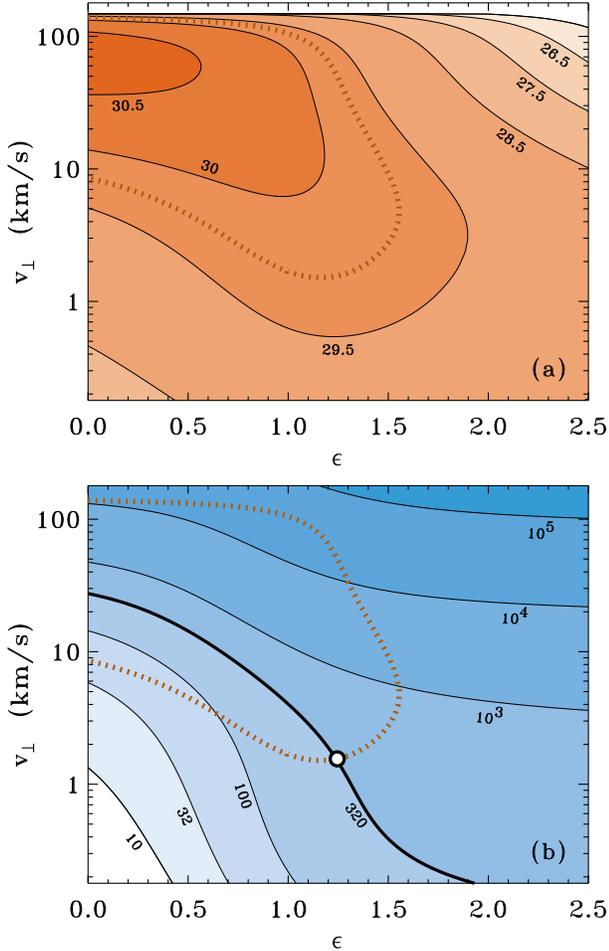}
\caption{Synthesized observations of (a) {\rm ROSAT} X-ray luminosity
and (b) ALMA 1.3~mm flux for AU~Mic, plotted as contours versus
modeled loop length distribution exponent $\varepsilon$ and
surface wave velocity amplitude $v_{\perp}$.
Contours are labeled with constant values of
(a) $\log_{10} L_{\rm X}$ (with $L_{\rm X}$ in units
of erg s$^{-1}$), and (b) $S_{\nu}$ in $\mu$Jy.
Models agreeing with the observed value of $L_{\rm X}$ are shown
in both panels with a dotted red contour, and the single model
that satisfies both
observational constraints is shown with a white circle.
\label{fig03}}
\end{figure}

Figure \ref{fig03}(a) shows contours of $L_{\rm X}$ computed for
the {\em ROSAT} PSPC band, and for a large grid of 250 $v_{\perp}$
values by 250 $\varepsilon$ values.\footnote{%
We chose not to include the smallest $v_{\perp}$ amplitudes in the
grid because there were no combinations of parameters in this
region that produced agreement with both the X-ray and millimeter
observations.  The $250 \times 250$ grid encompasses the ranges
$0.2 \leq v_{\perp} \leq 200$ km s$^{-1}$ and
$0 \leq \varepsilon \leq 2.5$.}
For any single value of $\varepsilon$, the X-ray luminosity first
increases with increasing $v_{\perp}$, because the temperatures
and densities in the loops are also increasing, but then $L_{\rm X}$
begins to decrease because the loops become too hot to produce
significant emission in the relatively soft PSPC passband.
The dotted red contour in Figure \ref{fig03}(a) denotes the
parameters that produce agreement with the observed value of
$\log L_{\rm X} = 29.74$ \citep{Hu99}.

The contours in Figure \ref{fig03}(b) show constant values of the
free-free emission flux $S_{\nu}$ computed at $\lambda = 1.3$ mm.
The thickest contour highlights the ALMA flux measured by
\citet{Mc13} for the central emission component of AU~Mic.
The corresponding observational contour for $L_{\rm X}$ is
reproduced from Figure \ref{fig03}(a), and it is noteworthy that
the two empirical constraints intersect one another (roughly
perpendicularly) at a single parameter value.
This best-fitting model has a basal velocity amplitude
$v_{\perp} = 1.5545$ km s$^{-1}$ and a loop distribution
power-law exponent $\varepsilon = 1.245$.
We note also that this model has a value of $v_{\perp}$ that falls
within the region of likely microturbulence values (1--2 km s$^{-1}$)
for early M-type dwarfs \citep{Gm82,Wd09,Bk11}.

The best-fitting model contains loops with peak temperatures between
3.7 MK (for $L_{\rm min}$) and 7.2 MK (for $L_{\rm max}$).
\citet{SS10} produced a fit of the full {\em Chandra} spectrum of
AU~Mic with three plasma components having temperatures of 3.37,
7.78, and 17.3 MK.
Our range of modeled $T_{\rm max}$ values is reasonably consistent
with the first two {\em Chandra} components.
The hottest observed X-ray component may be related to intermittent
nonthermal stellar flaring, which we did not include in our model.

The best-fitting model also exhibits loop-top electron densities 
between $7 \times 10^{11}$ cm$^{-3}$ (for $L_{\rm min}$) and
$3 \times 10^{7}$ cm$^{-3}$ (for $L_{\rm max}$), with the
densities increasing as one goes down from the loop tops.
\citet{Te04} found upper limits on electron densities from
{\em Chandra} spectroscopy of AU~Mic that are close to the above
values.
For lines of \ion{O}{7} and \ion{Mg}{11}, formed at roughly
2 MK and 7 MK respectively, \citet{Te04} found upper limits of
$5.6 \times 10^{11}$ and $5.6 \times 10^{12}$ cm$^{-3}$.
We can also combine our model densities with the $T_{\rm max}$
values given above to determine the range of gas pressures in
the loops and compare them with observations.
We assumed that $P$ remained constant over the length of each loop,
and the values spanned from 760 dyne cm$^{-2}$ (for $L_{\rm min}$) to
1.8 dyne cm$^{-2}$ (for $L_{\rm max}$).
\citet{Dz02} found a best fit for the UV-derived differential
emission measure of AU~Mic to occur for a constant pressure
of $\sim$3 dyne cm$^{-2}$, which falls inside the range of our
modeled loops.

\begin{figure}
\epsscale{1.17}
\plotone{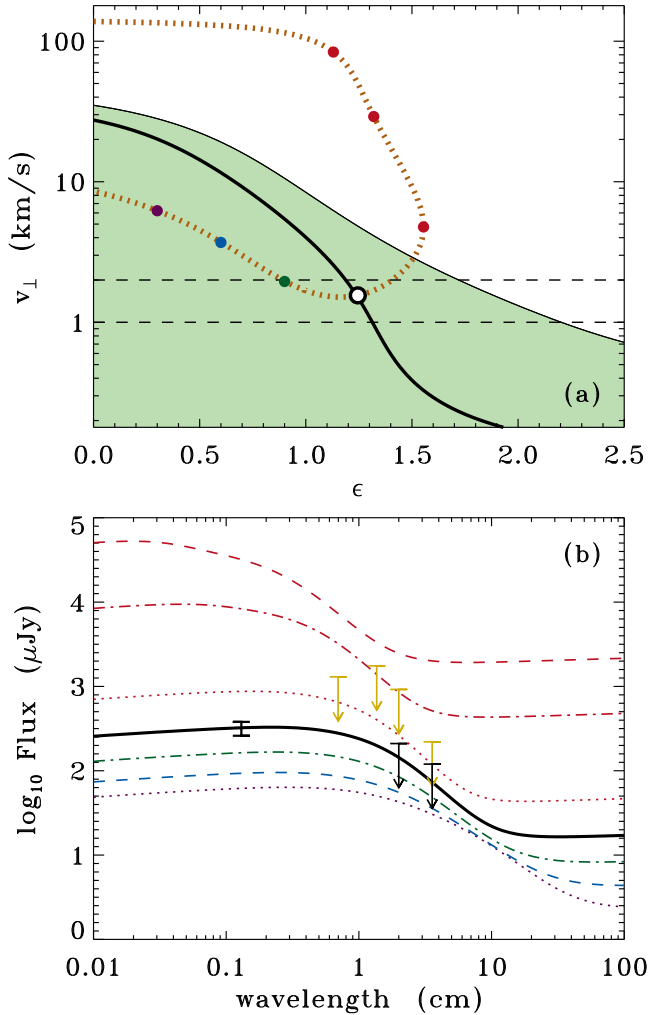}
\caption{(a) Contours showing models that agree with observed
$L_{\rm X}$ (dotted red curve) and ALMA 1.3~mm flux
(thick black curve) for
AU Mic, together with the region of parameter space that is
consistent with the \citet{Wh94} radio upper limit at
$\lambda = 2$ cm (green area).
Expected values of $v_{\perp}$ from M-dwarf microturbulence
are shown with thin black dashed lines.
(b) Synthesized radio/mm spectra for models denoted by
circles in panel (a).
From bottom to top, the spectra correspond to models that follow
the empirical $L_{\rm X}$ contour counterclockwise from
bottom-left to top-middle in panel (a).
Also shown are upper limits from \citet{Wh94} (black arrows) and
\citet{Lt00} (gold arrows), and the \citet{Mc13} 1.3~mm
measurement (black strut).
\label{fig04}}
\end{figure}

We also computed the free-free emission at longer radio wavelengths,
for which there are no firm detections of AU~Mic in quiescence.
Both \citet{Wh94} and \citet{Lt00} gave non-detection upper limits
that we can use as further tests of our model.
Figure \ref{fig04}(a) carries over the two empirical contours
from Figure \ref{fig03} and also divides the parameter space into
two regions: one in which the modeled emission is consistent with
the \citet{Wh94} upper limit for $\lambda = 2$ cm,
and one in which the modeled emission exceeds that limit.
Our best-fitting solution sits comfortably in the region that
is consistent with all of the observed radio upper limits.

Figure \ref{fig04}(b) shows representative radio and millimeter
spectra for a set of seven models that fall along the observational
$L_{\rm X}$ contour shown in Figure \ref{fig04}(a).
Each of these models reproduces the observed X-ray luminosity of
AU~Mic, but only one of them (coincidentally, the one with the
lowest value of $v_{\perp}$) agrees with the ALMA flux at 1.3~mm.
Four of these seven models have radio fluxes that fall below the
\citet{Wh94} and \citet{Lt00} upper limits.
It is noteworthy that a factor-of-two better measurement of the
radio continuum at 1--10 cm would provide a much more stringent
test of this model.

The shapes of the spectra in Figure \ref{fig04}(b) convey interesting
information about the spatial variations in temperature and density.
At the lowest wavelengths, the optical depth $\tau_{\nu}$ is small
(even when integrating all the way down to the coronal base), so
the $\nu^{2}$ dependence of the source function cancels out with
the $\nu^{-2}$ in the opacity.
Thus, the spectrum has a canonically flat shape that reflects only
the slow wavelength dependence of the Gaunt factor.
As the wavelength increases, the optical depth at the coronal base
gets larger, and more of the coronal loops become optically thick.
In that limit, the intensity is capped at a maximum value of
\begin{equation}
  I_{\nu} \, \approx \, 
  \frac{2 k_{\rm B} T \nu^2}{c^2}  \,\, .
  \label{eq:thick}
\end{equation}
If the temperature in the emitting region was a constant, this
would give a power-law spectrum going as $\lambda^{-2}$.
A steepening to this shape is indeed seen around
$\lambda \sim 1$--5 cm, but then it flattens out again.
This is because the optical depth keeps increasing as $\lambda$
increases, and the level of the $\tau_{\nu} = 1$ ``radio photosphere''
moves up from the coronal base to the tops of the loops.
Because $T(s)$ increases from the base to the loop-tops,
the emitting temperature to be used in Equation (\ref{eq:thick})
is an increasing function of $\lambda$, too.
This gives rise to an absolute value of $I_{\nu}$ that grows
progressively larger than it would have been had the emitting
temperature remained constant.
Presumably as $\lambda$ is increased even further, all of the loop
tops will become optically thick, the emergent free-free emission
will be dominated only by $T_{\rm max}$, and the spectrum will
start declining again as $\lambda^{-2}$.

Lastly, we mention that the resulting values of X-ray and radio/mm
emissivity do depend on the assumed value of the photospheric
magnetic filling factor $f_{\ast}$.
We ran a series of models for the standard parameter choices of
$v_{\perp} = 1.5545$ km s$^{-1}$ and $\varepsilon = 1.245$, where
$f_{\ast}$ was varied over two orders of magnitude from 0.01 to 1.
The X-ray luminosity scales roughly as
$L_{\rm X} \propto f_{\ast}^{1.26}$, and the free-free emission
flux at $\lambda = 1.3$ mm scales as
$S_{\nu} \propto f_{\ast}^{1.39}$.
These order-unity power law exponents imply that the uncertainties
in the coronal loop parameters determined above may be roughly of
the same magnitude as the uncertainty in our knowledge of $f_{\ast}$
for this star.

\section{Discussion and Conclusions}
\label{sec:conc}

In this paper we began the process of exploring whether a
physically motivated model of coronal loop heating could reproduce
the observed properties of AU~Mic.
We adapted an existing model of MHD turbulent dissipation to the
environment of an M1 dwarf star and treated the surface velocity
amplitude $v_{\perp}$ of stochastic Alfv\'{e}nic motions as a
free parameter.
For a value of $v_{\perp} \approx 1.5$ km s$^{-1}$ (which is
consistent with the expected granular microturbulence velocity)
and a power-law probability distribution of loop lengths $L$ that
goes roughly as $L^{-1.25}$, the model successfully predicts the
observed X-ray luminosity and millimeter-wave emission peak.
The model is also consistent with radio upper limits measured at
quiescent (non-flaring) time periods for AU~Mic.

It may be possible to use independent estimates of the stellar wind
mass loss rate of AU~Mic to say more about the verisimilitude of
the coronal heating model.
\citet{AB06} determined that a mass loss rate of order
$6 \times 10^{-12}$ $M_{\odot}$ yr$^{-1}$ for AU~Mic would generate
sufficient pressure to transport dust to the outermost parts of
its observed disk.
We used the stellar wind model presented by \citet{CS11} to compute
how the mass loss rate of AU~Mic depends on the assumed value of
$v_{\perp}$ in the treatment of coronal heating.
The ``default'' value of $v_{\perp} = 0.02$ km s$^{-1}$ obtained
from the convective turbulence model of \citet{Mz02} gives rise
to a mass loss rate of only
$2 \times 10^{-15} \, M_{\odot}$ yr$^{-1}$.
However, if we increase $v_{\perp}$ to the empirically determined
value of 1.5545 km s$^{-1}$ (see white circles in Figures
\ref{fig03}--\ref{fig04}), the \citet{CS11} mass loss rate
increases to $9 \times 10^{-12} \, M_{\odot}$ yr$^{-1}$.
This is close to the value required by \citet{AB06} to explain
the outer disk's dust diffusion.
It is at least circumstantial evidence that our loop model,
``calibrated'' with $v_{\perp} \approx 1.5$ km s$^{-1}$, reflects
what is actually going on in the outer atmosphere of AU~Mic.

Although our model accurately reproduces the magnitude of the
ALMA 1.3~mm central emission peak, we do not yet have a definitive
way to distinguish between this model and the idea of an inner
asteroid belt proposed by \citet{Mc13}.
For example, neither model predicts significant mid-infrared excess
centered on the star, and indeed none is seen for AU~Mic \citep{Li04}.
It is possible, of course, that AU~Mic has {\em both} an active
stellar corona and a bright inner source of dusty debris.
Additional information about the spectral energy distribution of
the central peak resolved by ALMA would be extremely helpful to
putting limits on the relative contributions of these two suggested
explanations for the 1.3~mm emission.
Also, better observational limits on the spatial extent of the
central peak would be helpful, since the dust belt is expected to
be several orders of magnitude larger in size than the source of
coronal emission.

In addition to better observational constraints, there are also
several ways that the models can be improved.
We limited ourselves to just the time-steady component of coronal
heating and not the more intermittent flaring, but the latter
is known to be a non-negligible component of the total X-ray
\citep{SS10} and ultraviolet \citep{Ro01} emission.
Because we did not model flares, we also ignored the potential
contributions of nonthermal electrons and gyromagnetic emission
to the radio and millimeter spectra \citep[see][]{Wh94,Li96}.
We did not simulate the full three-dimensional magnetic geometry
of the corona of AU~Mic; instead we treated its topological
complexity using a simple statistical distribution of loop sizes.
Future refinements to our predictions of radio, millimeter, and
X-ray emission from stellar coronae should include improvements
to each of the above approximations.

Finally, the success of any ab~initio coronal model depends on
our understanding of magnetic dynamos in low-mass stars.
Young, rapidly rotating stars are suspected of having more
vigorous convective motions and more frequent magnetic flux
emergence than older stars such as the Sun
\citep[e.g.,][]{Mn04,Ba07,Ka09,Br11}.
The turbulent convection models we have used so far \citep{Mz02} do
not yet include these effects, but our semi-empirical determination
of $v_{\perp} \approx 1.5$ km s$^{-1}$ may help put limits on
how active the surfaces of young stars actually are.
Still, we do not know what the granulation pattern---and the
distribution of magnetic flux tubes---really looks like on the
surface of a saturated-activity star.
It is also possible that rapid rotation strongly affects the
distribution of loop lengths \citep{Aa12}.
M dwarfs such as AU~Mic are also close to the dividing line
between having a radiative core and being fully convective.
The qualitative properties of rotation, dynamos, and large-scale
magnetic fields are believed to change substantially across this
dividing line \citep{MM01,RB07,Ir11,Gs13}.
These ``hidden'' aspects of stellar interiors may have significant
impacts on the high-energy activity of a star like AU~Mic and its
surrounding dust and debris.

\acknowledgments

The authors gratefully acknowledge Adriaan van Ballegooijen, Nancy
Brickhouse, and Hans Moritz G\"{u}nther for many valuable discussions.
This paper made use of the following ALMA data:
ADS/JAO.ALMA\#2011.0.00142.S.
ALMA is a partnership of ESO (representing its member states),
NSF (USA) and NINS (Japan), together with NRC (Canada) and NSC
and ASIAA (Taiwan), in cooperation with the Republic of Chile.
The Joint ALMA Observatory is operated by ESO, AUI/NRAO, and NAOJ.
The National Radio Astronomy Observatory is a facility of the
National Science Foundation operated under cooperative
agreement by Associated Universities, Inc.
This research made extensive use of NASA's Astrophysics
Data System and the SIMBAD database operated at CDS,
Strasbourg, France.
We also thank the anonymous referee for constructive suggestions
that improved this paper.

\end{document}